\documentclass[12pt]{article}
\usepackage{graphicx}
\usepackage{color}

\def\hybrid{\topmargin 0pt      \oddsidemargin 0pt
        \headheight 0pt \headsep 0pt
       \voffset-1cm
        \textwidth 6.25in       
       \textheight 9.5in       
        \marginparwidth 0.0in
        \parskip 5pt plus 1pt   \jot = 1.5ex}
\catcode`\@=11
\def\marginnote#1{}

\newcount\hour
\newcount\minute
\newtoks\amorpm
\hour=\time\divide\hour by60
\minute=\time{\multiply\hour by60 \global\advance\minute by-\hour}
\edef\standardtime{{\ifnum\hour<12 \global\amorpm={am}%
        \else\global\amorpm={pm}\advance\hour by-12 \fi
        \ifnum\hour=0 \hour=12 \fi
        \number\hour:\ifnum\minute<10 0\fi\number\minute\the\amorpm}}
\edef\militarytime{\number\hour:\ifnum\minute<10 0\fi\number\minute}

\def\draftlabel#1{{\@bsphack\if@filesw {\let\thepage\relax
   \xdef\@gtempa{\write\@auxout{\string
      \newlabel{#1}{{\@currentlabel}{\thepage}}}}}\@gtempa
   \if@nobreak \ifvmode\nobreak\fi\fi\fi\@esphack}
        \gdef\@eqnlabel{#1}}
\def\@eqnlabel{}
\def\@vacuum{}
\def\draftmarginnote#1{\marginpar{\raggedright\scriptsize\tt#1}}

\def\draftlabel#1{{\@bsphack\if@filesw {\let\thepage\relax
   \xdef\@gtempa{\write\@auxout{\string
      \newlabel{#1}{{\@currentlabel}{\thepage}}}}}\@gtempa
   \if@nobreak \ifvmode\nobreak\fi\fi\fi\@esphack}
        \gdef\@eqnlabel{#1}}
\def\@eqnlabel{}
\def\@vacuum{}
\def\draftmarginnote#1{\marginpar{\raggedright\scriptsize\tt#1}}

\def\draft{\oddsidemargin -.5truein
        \def\@oddfoot{\sl preliminary draft \hfil
        \rm\thepage\hfil\sl\today\quad\militarytime}
        \let\@evenfoot\@oddfoot \overfullrule 3pt
        \let\label=\draftlabel
        \let\marginnote=\draftmarginnote
   \def\@eqnnum{(\theequation)\rlap{\kern\marginparsep\tt\@eqnlabel}%
\global\let\@eqnlabel\@vacuum}  }

\def\numberbysection{\@addtoreset{equation}{section}
        \def\theequation{\thesection.\arabic{equation}}}

\def\underline#1{\relax\ifmmode\@@underline#1\else
        $\@@underline{\hbox{#1}}$\relax\fi}

\def\titlepage{\@restonecolfalse\if@twocolumn\@restonecoltrue\onecolumn
     \else \newpage \fi \thispagestyle{empty}\c@page\z@
        \def\thefootnote{\fnsymbol{footnote}} }

\def\endtitlepage{\if@restonecol\twocolumn \else  \fi
        \def\thefootnote{\arabic{footnote}}
        \setcounter{footnote}{0}}  
\relax

\hybrid

\newfont{\Bbb}{msbm10 scaled 1\@ptsize00}
\newfont{\Bbbb}{msbm7 scaled 1\@ptsize00}
\newcommand{\CC}{\mbox{\Bbb C}}

\newcommand{\DDD}{\raise-1pt\hbox{$\mbox{\Bbbb D}$}}

\newcommand{\UUU}{\raise-1pt\hbox{$\mbox{\Bbbb U}$}}

\newcommand{\ZZ}{\mbox{\Bbb Z}}
\newcommand{\z}{\raise-1pt\hbox{$\mbox{\Bbbb Z}$}}
\newcommand{\ka}{\kappa}

\def\beq{\begin{equation}}
\def\eeq{\end{equation}}
\def\p{\partial}

\begin{document}

\begin{titlepage}

\title{KZ-Calogero correspondence revisited}

\author{A.~Zabrodin\thanks{National Research University Higher School of Economics,
Russian Federation; Institute of Biochemical Physics of Russian Academy of Sciences,
Kosygina str. 4, 119334, Moscow, Russia;
e-mail: zabrodin@itep.ru}
\and
 A. Zotov\thanks{Steklov Mathematical
Institute of Russian Academy of Sciences, Gubkina str. 8, 119991,
Moscow, Russia; ITEP, 25 B.Cheremushkinskaya, Moscow 117218, Russia;
Moscow Institute of Physics and Technology, Inststitutskii per.  9,
Dolgoprudny, Moscow region, 141700, Russia; e-mail: zotov@mi.ras.ru}
}

\date{January 2017}
\maketitle

\vspace{-7cm} \centerline{ \hfill ITEP-TH-03/17}\vspace{7cm}

\begin{abstract}

We discuss the correspondence between the Knizhnik-Zamolodchikov equations
associated with $GL(N)$ and the $n$-particle quantum Calogero model in the case
when $n$ is not necessarily equal to $N$.
This can be viewed
as a natural ``quantization'' of the quantum-classical correspondence
between quantum Gaudin and classical Calogero models.

\end{abstract}

\end{titlepage}

\vspace{5mm}

\section{Introduction}

The rational Knizhnik-Zamolodchikov (KZ) equations \cite{KZ} have the form
\beq\label{kz1}
\hbar \p_{x_i}\Bigl | \Phi \Bigr >=\left ({\bf g}^{(i)}+
\kappa \sum_{j\neq i}^n \frac{{\bf P}_{ij}}{x_i-x_j}\right )
\Bigl | \Phi \Bigr >
\eeq
where $\Bigl | \Phi \Bigr >=\Bigl | \Phi \Bigr >(x_1, \ldots , x_n)$ belongs to the
tensor product ${\cal V}=V\otimes V\otimes \ldots \otimes V =V^{\otimes n}$ of the
vector spaces $V=\CC^N$, ${\bf P}_{ij}$ is the permutation of the $i$-th and $j$-th
factors, ${\bf g}=\mbox{diag}(g_1, \ldots , g_N)$ is a diagonal $N\! \times \! N$ matrix and
${\bf g}^{(i)}$ is the operator in ${\cal V}$ acting as ${\bf g}$ on the $i$-th factor
(and identically on all other factors).

The remarkable correspondence of the KZ equations with the
quantum Calogero model \cite{Calogero} defined by the Hamiltonian
\beq\label{cal1}
\hat {\cal H}=\hbar^2\sum_{i=1}^n \p_{x_i}^2 -
\sum_{i\neq j}^n \frac{\kappa (\kappa \! -\! \hbar )}{(x_i-x_j)^2}
\eeq
was established by Matsuo and Cherednik
in \cite{Matsuo,Cherednik} (see also \cite{FV,Veselov}) in the case
$N=n$. In this case one can find solutions to (\ref{kz1}) in the form
$$
\Bigl | \Phi \Bigr >=\sum_{\sigma \in S_n}\Phi_{\sigma}\Bigl | e_{\sigma} \Bigr >, \qquad
\Bigl | e_{\sigma} \Bigr >=e_{\sigma (1)}\otimes e_{\sigma (2)}\otimes \ldots \otimes
e_{\sigma (n)},
$$
where $e_{a}$ are standard basis vectors in $V=\CC^n=\CC^N$ and $S_n$ is the symmetric group.
If such $\Bigl | \Phi \Bigr >$ solves the KZ equations,
then the function
\beq\label{cal2}
\Psi =\sum_{\sigma \in S_n}\Phi_{\sigma}
\eeq
is an eigenfunction of the Calogero Hamiltonian:
\beq
\label{cal3}
\hat {\cal H}\Psi =E\Psi , \qquad E=g_1^2 +g_2^2 +\ldots +g_N^2.
\eeq
This correspondence can be extended to the trigonometric versions of the both models.

We will show that a similar correspondence exists also in the case
when the number of marked points $n$ is not necessarily equal to
$N=\mbox{dim}\, V$. In this form it looks like a quantum deformation
of the quantum-classical correspondence
\cite{AKLTZ13,ALTZ14,GZZ14,MTV12,Z2} between the quantum Gaudin
and classical Calogero models (see \cite{BG} for a discussion of the 
Matsuo-Cherednik map in this context).

The system of KZ equations is a non-stationary version of the quantum Gaudin model, with
$\hbar$ being the parameter of non-stationarity. We denote it as $\hbar$ because it becomes
the true Planck constant in the corresponding quantum Calogero model.
The spectral problem for the Gaudin model is a ``quasiclassical'' limit
of KZ as $\hbar \to 0$. Indeed, as $\hbar \to 0$ the
KZ solutions have the asymptotic form \cite{RV}
$$
\Bigl |\Phi \Bigr >=\left ( \Bigl |\phi_0\Bigr > + \hbar \Bigl |\phi_1\Bigr >
+\ldots \right )
e^{S/\hbar}
$$
which, upon substitution to the KZ equations (\ref{kz1}), leads, in the leading order,
to the
joint eigenvalue problems
$$
{\bf H}_i \Bigl |\phi_0\Bigr >=p_i \Bigl |\phi_0\Bigr >, \qquad
p_i =\frac{\p S}{\p x_i},\qquad i=1, \ldots , n,
$$
for the commuting Gaudin Hamiltonians
$\displaystyle{
{\bf H}_i = {\bf g}^{(i)}+
\kappa \sum_{j\neq i}^n \frac{{\bf P}_{ij}}{x_i-x_j}}
$
with the Planck constant $\kappa$.
In the quan\-tum-clas\-si\-cal correspondence, the eigenvalues $p_i$ are identified with
momenta of the Calogero-Moser particles with coordinates $x_i$.

The plan of the paper is as follows. In section 2, we describe the rational Gaudin model
with a formal Planck constant $\kappa$ and the associated KZ equations. In section 3
the KZ-Calogero correspondence is established. Section 4 is devoted to the trigonometric
version of the correspondence. Finally, in section 5 we discuss the interpretation of the results
as a ``quantum'' deformation of the quantum-classical correspondence. Section 6 is the conclusion.
In the appendix we show that the wave function from section 3 is also an eigenfunction of the higher
Calogero Hamiltonian $\hat {\cal H}_3$.

\section{The Gaudin Hamiltonians and KZ equations}

Let ${\sf e}_{ab}^{\kappa}$ be generators of the
``$\kappa$-dependent version'' of the universal
enveloping algebra
$U(gl(N))$ with the commutation relations
$[{\sf e}^{\kappa}_{ab}, \,{\sf e}^{\kappa}_{a'b'}]=
{\kappa}(\delta_{a'b}{\sf e}^{\kappa}_{ab'}-
\delta_{ab'}{\sf e}^{\kappa}_{a'b})$. Since at $\kappa =0$
the operators ${\sf e}_{ab}^{\kappa}$ commute,
the parameter $\kappa$ plays the role of the formal Planck's constant.
Let $\pi$ be the $N$-dimensional vector representation
of $U^{(\kappa )}(gl(N))$.
We have $\pi ({\sf e}^{\kappa}_{ab})=\kappa e_{ab}$, where
$e_{ab}$ is the standard
basis in the space of $N\! \times \! N$ matrices:
the matrix $e_{ab}$ has only
one non-zero element (equal to 1) at the place $ab$:
$(e_{ab})_{a'b'}=\delta_{aa'}\delta_{bb'}$.
Note that ${\bf I}=\sum_a e_{aa}$ is the unity operator and
${\bf P}=\sum_{ab}e_{ab} \otimes e_{ba}$ is the permutation operator
acting in the space $\CC^N \otimes \CC^N$.

In the tensor product $U^{(\kappa )}(gl(N))^{\otimes n}$
the generators
${\sf e}^{\kappa}_{ab}$
can be realized as
${\sf e}^{\kappa \, (i)}_{ab}:= {\bf I}^{\otimes (i-1)} \otimes
{\sf e}^{\kappa}_{ab} \otimes  {\bf I}^{\otimes (n-i)}$.
It is clear that they commute for  any $i \ne j$ and any $a,b$
because act non-trivially in different spaces.
Similarly, for any matrix ${\bf g}\in \mbox{End}(\CC^N)$ we
define ${\bf g}^{(i)}$ acting in the tensor product
${\cal V}=(\CC^N)^{\otimes n}$:
${\bf g}^{(i)}= {\bf I}^{\otimes (i-1)} \otimes
{\bf g}\otimes  {\bf I}^{\otimes (n-i)}\in \mbox{End}({\cal V})$.
In this notation,
$\displaystyle{{\bf P}_{ij}:=\sum_{a,b}e^{(i)}_{ab}e^{(j)}_{ba}}$ is the
permutation operator of the $i$-th and $j$-th tensor factors
in ${\cal V}=\CC^N \otimes \ldots \otimes \CC^N$. Clearly, ${\bf P}_{ij}={\bf P}_{ji}$ and
${\bf P}^2_{ij}={\bf I}$.

Fix $n$ distinct numbers $x_i\in \CC$ and
a diagonal $N\! \times \! N$
matrix ${\bf g} =\mbox{diag}\,(g_1, \ldots , g_N)$.
(We assume that $n\geq N$ and that the $g_i$'s are all distinct and non-zero.)
We will call ${\bf g}$ the twist matrix.
The commuting Gaudin Hamiltonians are
\beq\label{gaudin-gen}
{\bf H}_i=\frac{1}{\kappa}\left (\sum_{a=1}^{N}g_a {\sf e}^{\kappa \, (i)}_{aa} +
\sum_{j\neq i}\sum_{a,b=1}^{N}
\frac{{\sf e}^{\kappa \,(i)}_{ab}{\sf e}^{\kappa \,(j)}_{ba}}{x_i-x_j}\right )\,,
\quad i=1,\ldots , n.
\eeq
The Hamiltonians of the quantum
Gaudin model \cite{Gaudin} with the Hilbert space
${\cal V}=(\CC^N)^{\otimes n}$ are restrictions of the operators
(\ref{gaudin-gen}) to the $N$-dimensional vector representation $\pi$:
\begin{equation}\label{D4}
{\bf H}_i= \sum_{a=1}^{N}g_a e^{(i)}_{aa} +
\kappa \sum_{j\neq i}\sum_{a,b=1}^{N}
\frac{e^{(i)}_{ab}e^{(j)}_{ba}}{x_i-x_j}=
 {\bf g} ^{(i)} + \kappa \sum_{j\neq i}\frac{{\bf P}_{ij}}{x_i-x_j}
\,, \quad i=1,\ldots , n
\end{equation}
(for brevity we denote $\pi ^{\otimes n}({\bf H}_i)$ by the same letter
${\bf H}_i$). It is known that the Gaudin Hamiltonians form a commutative family:
$[{\bf H}_i , {\bf H}_j]=0$ for all $i,j =1, \ldots , n$.

The operators
\beq\label{Ma}
{\bf M}_a=\sum_{l=1}^{n}e_{aa}^{(l)}
\eeq
commute among themselves and with the
Gaudin Hamiltonians: $[{\bf H}_i, {\bf M}_a]=0$.
Clearly,
$\sum_a {\bf M}_a =n {\bf I}$, and
$
\displaystyle{\sum_{i=1}^{n}{\bf H}_i= \sum_{a=1}^{N}g_a {\bf M}_a}
$.
The joint spectral problem is
$$
\left \{\begin{array}{l}
{\bf H}_i \Bigl |\phi \Bigr > =H_i \Bigl |\phi \Bigr >
\\
{\bf M}_a \Bigl |\phi \Bigr > =M_a \Bigl |\phi \Bigr >
\end{array}
\right.
$$
The common eigenstates of the Hamiltonians can be classified
according to eigenvalues of the operators ${\bf M}_a$.

Let
\beq\label{weight1}\displaystyle{
{\cal V}=V^{\otimes n} \,\, =
\bigoplus_{M_1, \ldots , M_N} \!\! \!\! {\cal V}(\{M_a \})}
\eeq
be the weight decomposition of the Hilbert space ${\cal V}$ of the Gaudin model
into the direct sum of eigenspaces for the operators
${\bf M}_a$ with the eigenvalues $M_a \in \ZZ_{\geq 0}$, $a=1, \ldots , N$
(recall that $M_1 +\ldots +M_N=n$).
Then the eigenstates of ${\bf H}_i$'s are in the spaces ${\cal V}(\{M_a \})$,
$$
\mbox{dim} {\cal V}(\{M_a \})=\frac{n!}{M_1! \ldots M_N!}.
$$
The basis vectors in ${\cal V}(\{M_a \})$ are
$
\Bigl |J\Bigr > =e_{j_1}\otimes e_{j_2}\otimes \ldots \otimes e_{j_n},
$
where the number of indices $j_k$ such that $j_k =a$ is equal to $M_a$ for all
$a=1, \ldots , N$.
We also introduce dual vectors $\Bigl <J\Bigr |=
e_{j_1}^{\dag}\otimes e_{j_2}^{\dag}\otimes \ldots \otimes e_{j_n}^{\dag}$ such that
$\Bigl <J\Bigr |J'\Bigr >=\delta_{J, J'}$.

The system of KZ equations is a non-stationary version of the Gaudin model:
\beq\label{na1}
\hbar \p_{x_i}\Bigl |\Phi \Bigr >={\bf H}_i \Bigl |\Phi \Bigr >, \qquad i=1, \ldots , n.
\eeq
It respects the weight decomposition (\ref{weight1}), hence the solutions
belong to the weight subspaces ${\cal V}(\{M_a \})$. Equations (\ref{na1})
are compatible due to the flatness conditions
\beq\label{na1a}
[\hbar \p_{x_i}-{\bf H}_i, \, \hbar \p_{x_j}-{\bf H}_j]=0 \qquad \mbox{for all $\,\,\, i,j=1, \ldots , n$}.
\eeq

\section{The KZ-Calogero correspondence}

We claim that for any solution of the KZ equations belonging to the space
${\cal V}(\{M_a \})$,
$\displaystyle{\Bigl |\Phi \Bigr >=\sum_J \Phi_J \Bigl |J\Bigr >}$,
the function
\beq\label{kz2}
\Psi =\sum_J \Phi_J
\eeq
is an eigenfunction of the Calogero Hamiltonian with the eigenvalue
$\displaystyle{E=\sum_{a=1}^N M_a g_a^2}$:
\beq\label{kz3}
\left ( \hbar^2 \sum_{i=1}^n \p_{x_i}^2-\sum_{i\neq j}^n
\frac{\kappa (\kappa -\hbar )}{(x_i-x_j)^2}\right )\Psi = E\Psi\,.
\eeq
In particular, at $n=N$ and $M_1=M_2=\ldots =M_N=1$ we get the result of
\cite{Matsuo,Cherednik,FV,Veselov}.

For the proof consider the covector equal to the sum of all basis (dual) vectors
from the space $({\cal V}(\{M_a\}))^*$:
$$
\Bigl < \Omega \Bigr |=\sum_J \Bigl < J \Bigr |,
$$
then $\Psi =\Bigl <\Omega \Bigr | \Phi \Bigr >$.
Applying the operator $\hbar \p_{x_i}$ to the KZ equation (\ref{kz1}), we get:
$$
\hbar^2 \p_{x_i}^2 \Bigl |\Phi \Bigr >=
-\hbar \kappa \sum_{j\neq i}\frac{{\bf P}_{ij}}{(x_i-x_j)^2}\Bigl |\Phi \Bigr >+
\left ({\bf g}^{(i)}+\kappa \sum_{j\neq i}\frac{{\bf P}_{ij}}{x_i-x_j}\right )
\hbar \p_{x_i}\Bigl |\Phi \Bigr >
$$
$$
=\,
-\hbar \kappa \sum_{j\neq i}\frac{{\bf P}_{ij}}{(x_i-x_j)^2}\Bigl |\Phi \Bigr >+
\left ({\bf g}^{(i)}+\kappa \sum_{j\neq i}\frac{{\bf P}_{ij}}{x_i-x_j}\right )
\left ({\bf g}^{(i)}+\kappa \sum_{l\neq i}\frac{{\bf P}_{il}}{x_i-x_l}\right )
\Bigl |\Phi \Bigr >
$$
$$
=\,
-\hbar \kappa \sum_{j\neq i}\frac{{\bf P}_{ij}}{(x_i-x_j)^2}\Bigl |\Phi \Bigr >+
\kappa^2 \sum_{j\neq i}\frac{1}{(x_i-x_j)^2}\Bigl |\Phi \Bigr >+({\bf g}^{(i)})^2\Bigl |\Phi \Bigr >
$$
$$
+\, \kappa^2 \sum_{j\neq l\neq i}\frac{{\bf P}_{ij}{\bf P}_{il}}{(x_i-x_j)(x_i-x_l)}\Bigl |\Phi \Bigr >
+\kappa \sum_{j\neq i}\frac{{\bf P}_{ij}}{x_i-x_j}{\bf g}^{(i)}\Bigl |\Phi \Bigr >+
\kappa \sum_{j\neq i}\frac{{\bf P}_{ij}}{x_i-x_j}{\bf g}^{(j)}\Bigl |\Phi \Bigr >.
$$
In the last lines we took into account that ${\bf P}_{ij}^2={\bf I}$
and ${\bf g}^{(i)}{\bf P}_{ij}={\bf P}_{ij}{\bf g}^{(j)}$.
Since $\Bigl <\Omega \Bigr |{\bf P}_{ij}\Bigl |J\Bigr >=1$ for all
basis vectors $\Bigl |J\Bigr >$,
we have
$\Bigl <\Omega \Bigr |{\bf P}_{ij}=\Bigl <\Omega \Bigr |$.
Therefore, the
permutation operators disappear after applying $\Bigl <\Omega \Bigr
|$ from the left. Summing $\hbar^2 \Bigl <\Omega \Bigr |\p_{x_i}^2
\Bigl |\Phi \Bigr >=\hbar^2 \p_{x_i}^2 \Psi$ over $i$ and using the
identities\footnote{Identity (\ref{kz4}) follows from
$\frac{1}{(x_i-x_j)}\frac{1}{(x_i-x_l)}+\frac{1}{(x_i-x_j)}\frac{1}{(x_l-x_j)}+\frac{1}{(x_i-x_l)}\frac{1}{(x_j-x_l)}=0$
 applied to the sum symmetrized with respect to $i,j,l$.}
 \beq\label{kz4}
\sum_{j\neq l\neq i}\frac{1}{(x_i-x_j)(x_i-x_l)}=0,
\eeq
\beq\label{kz5}
\sum_{i\neq j}\frac{{\bf g}^{(i)}+{\bf g}^{(j)}}{x_i-x_j}=0,
\eeq
\beq\label{kz6}
\sum_i\Bigl <\Omega \Bigr |({\bf g}^{(i)})^2\Bigl |\Phi \Bigr >=\left (
\sum_{a=1}^N M_ag_a^2\right )\Psi
\eeq
we get (\ref{kz3}).

Note that $\Psi =\Bigl <\Omega \Bigr | \Phi \Bigr >$ is an
eigenfunction of the total momentum operator $\hat {\cal P}=\hbar
\sum_{j}\p_{x_j}$ with the eigenvalue $\sum_a M_a g_a$. In the appendix it is shown that
$\Psi$ is also an eigenfunction of the cubic Calogero Hamiltonian $\hat {\cal H}_3$.
We conjecture that
$\Psi$ is the common eigenfunction for all
higher Calogero Hamiltonians $\hat {\cal H}_k$ with the eigenvalues
$E_k=\sum_a M_a g_a^k$. The first four Hamiltonians are explicitly written in \cite{UWH}.

\section{Trigonometric case}

The trigonometric (hyperbolic) version of the system of KZ equations reads \cite{FV}
\beq\label{trig1}
\hbar \p_{x_i}\Bigl | \Phi \Bigr >=\left ({\bf g}^{(i)}+
\kappa \gamma \sum_{j\neq i}^n \Bigl ( \coth \gamma (x_i-x_j){\bf P}_{ij}+{\bf T}_{ij}\Bigr )\right )
\Bigl | \Phi \Bigr >,
\eeq
where we use the same notation as in (\ref{kz1}) and
$$
{\bf T}=\sum_{a>b}(e_{ab}\otimes e_{ba} - e_{ba}\otimes e_{ab}).
$$
This operator acts on basis vectors as follows:
\beq\label{trig2}
{\bf T}e_a \otimes e_b =\left \{
\begin{array}{l} \phantom{-}
e_b \otimes e_a \qquad \mbox{if} \quad a<b
\\ -\, e_b \otimes e_a \qquad \mbox{if} \quad a>b
\\ \phantom{-\, }0 \phantom{\otimes e_a} \qquad \!\! \quad\mbox{otherwise}
\end{array}\right.
\eeq
Note that ${\bf T}_{ji}=-{\bf T}_{ij}$.
In the limit $\gamma \to 0$ we recover the rational KZ equations (\ref{kz1}).

A calculation similar to the one given above in the rational setting leads to
the following statement.
For any solution of the KZ equations (\ref{trig1}) belonging to the space
${\cal V}(\{M_a \})$,
$\Bigl |\Phi \Bigr >=\sum_J \Phi_J \Bigl |J\Bigr >$,
the function
$\Psi =\sum_J \Phi_J$
solves the spectral problem for the Calogero-Sutherland Hamiltonian
\beq\label{trig4}
\left ( \hbar^2 \sum_{i=1}^n \p_{x_i}^2-\sum_{i\neq j}^n
\frac{\kappa (\kappa -\hbar )\gamma^2}{\sinh ^2\gamma(x_i-x_j)}\right )\Psi = E\Psi
\eeq
with the eigenvalue
\beq\label{trig3}
E=\sum_{a=1}^N M_a g_a^2 +\frac{\kappa^2\gamma^2}{3}\sum_{a=1}^N M_a(M_a^2-1).
\eeq

Here are some details of the calculation which is actually more involved
than in the rational case.
Applying the operator $\hbar \p_{x_i}$ to the KZ equation (\ref{trig1}), we get:
$$
\hbar^2 \p_{x_i}^2 \Bigl |\Phi \Bigr >=-\,
\hbar \kappa \gamma^2 \sum_{j\neq i}\frac{{\bf P}_{ij}}{\sinh^2 \gamma (x_i-x_j)} \, \Bigl | \Phi \Bigr >
$$
$$
+
\left ({\bf g}^{(i)}+
\kappa \gamma \!
\sum_{j\neq i}^n \Bigl ( \coth \gamma (x_i\! -\! x_j){\bf P}_{ij}+{\bf T}_{ij}\Bigr )\right )
\left ({\bf g}^{(i)}+
\kappa \gamma \!
\sum_{l\neq i}^n \Bigl ( \coth \gamma (x_i\!-\! x_l){\bf P}_{il}+{\bf T}_{il}\Bigr )\right )\!\!
\Bigl | \Phi \Bigr >.
$$
Again, in order to obtain an equation for $\Psi =\Bigl <\Omega \Bigr |\Phi \Bigr >$
we apply $\Bigl <\Omega \Bigr |=\sum_J \Bigl <J\Bigr |$ from the left and sum over $i$.
After opening brackets in the right hand side several different terms appear,
``wanted'' and ``unwanted'' ones. The ``wanted'' terms
are
$$
-\hbar \kappa \gamma^2 \sum_{j\neq i}\frac{1}{\sinh^2 \gamma (x_i-x_j)}\Psi +
\kappa^2\gamma^2 \sum_{i\neq j}\coth ^2 \gamma (x_i\! -\! x_j)\Psi
$$
$$
=\, \sum_{i\neq j}^n
\frac{\kappa (\kappa -\hbar )\gamma^2}{\sinh ^2\gamma(x_i-x_j)}\Psi
+n(n-1) \kappa^2 \gamma^2\Psi .
$$
It appears that the ``unwanted'' terms either cancel or contribute to the eigenvalue. To see this,
we need some identities. First of all, the trigonometric analog of identity (\ref{kz4})
 is\footnote{Similarly to (\ref{kz4}) identity (\ref{trig5}) follows
 from the summation formula for $\coth$ function:
$\coth\gamma{(x_i-x_j)}\coth\gamma{(x_i-x_l)}+\coth\gamma{(x_i-x_j)}\coth\gamma{(x_l-x_j)}+\coth\gamma{(x_i-x_l)}\coth\gamma{(x_j-x_l)}=1$
 and $\sum_{i\neq j\neq l} 1=n(n-1)(n-2)$.}
 \beq\label{trig5}
\sum_{i\neq j\neq l}\coth \gamma (x_i-x_j)\coth \gamma (x_i-x_l)=
\frac{1}{3}\, n(n-1)(n-2).
\eeq
An obvious trigonometric analog of (\ref{kz5}) is
\beq\label{trig5a}
\sum_{i\neq j}\coth \gamma (x_i-x_j)({\bf g}^{(i)}+{\bf g}^{(j)})=0.
\eeq
Using (\ref{trig2}), one can prove the identity
\beq\label{trig6}
({\bf g}^{(i)}-{\bf g}^{(j)}){\bf T}_{ij}+{\bf T}_{ij}
({\bf g}^{(i)}-{\bf g}^{(j)})=0.
\eeq
 The most non-trivial identities are
\beq\label{trig7}
\sum_{i\neq j}\Bigl <\Omega \Bigr |{\bf T}^2_{ij}\Bigl |\Phi \Bigr >=-\left (
n(n-1)-\sum_a M_a(M_a-1)\right )\Psi ,
\eeq
\beq\label{trig8}
\sum_{i\neq j\neq l}\Bigl <\Omega \Bigr |{\bf T}_{ij}{\bf T}_{il}\Bigl |\Phi \Bigr >=
-\frac{1}{3}\left (n(n-1)(n-2)-\sum_a M_a (M_a-1)(M_a-2)\right )\Psi .
\eeq
 They are derived from the definition (\ref{trig2}). Consider first
 (\ref{trig7}). The operator ${\bf T}^2_{ij}$ acts on arbitrary $e_a^{(i)}
 e_b^{(j)}$ entering $|\Phi \Bigr >\in {\cal V}(\{M_a \})$ as follows:
 \beq\label{trig81}
{\bf T}^2_{ij}\ e_a^{(i)}
 e_b^{(j)} =\left \{
\begin{array}{l} -
e_a^{(i)} e_b^{(j)} \qquad \mbox{if} \quad a\neq b
\\ \phantom{-\, }0 \phantom{\otimes e_a} \qquad \!\!
\quad\mbox{otherwise}.
\end{array}\right.
 \eeq
 Therefore, we compute $\sum_{i\neq j}1=n(n-1)$ for all $|J \Bigr >$ and
 subtract the terms corresponding to the second line of
 (\ref{trig81}). To prove (\ref{trig8}), it is convenient to
 symmetrize ${\bf T}_{ij}{\bf T}_{il}$ with respect to permutations of
 $i,j,l$ (keeping in mind that ${\bf T}_{ij}=-{\bf T}_{ji}$):
 $$
\sum_{i\neq j\neq l}\Bigl <\Omega \Bigr |{\bf T}_{ij}{\bf
T}_{il}\Bigl |\Phi \Bigr >=\frac{1}{3}\sum_{i\neq j\neq l}\Bigl
<\Omega \Bigr |{\bf T}_{ij}{\bf T}_{il}+{\bf T}_{lj}{\bf
T}_{ij}+{\bf T}_{il}{\bf T}_{jl}\Bigl |\Phi \Bigr >
 $$
 It can be verified directly that the operator ${\bf T}_{ij}{\bf T}_{il}+{\bf T}_{lj}{\bf
T}_{ij}+{\bf T}_{il}{\bf T}_{jl}$ acts on arbitrary $e_a^{(i)}
 e_b^{(j)} e_c^{(l)}$ entering $|\Phi \Bigr >\in {\cal V}(\{M_a \})$ as follows:
 \beq\label{trig82}
\left({\bf T}_{ij}{\bf T}_{il}+{\bf T}_{lj}{\bf T}_{ij}+{\bf
T}_{il}{\bf T}_{jl}\right)\ e_a^{(i)}
 e_b^{(j)} e_c^{(l)} =\left \{
\begin{array}{l}
 \phantom{-\, }0 \qquad\quad \mbox{if} \ a=b=c
\\ - e_a^{(l)}
 e_b^{(i)} e_c^{(j)} \ \
\mbox{otherwise}.
\end{array}\right.
 \eeq
 Therefore, we again compute $\sum_{i\neq j\neq l}1=n(n-1)(n-2)$,
 then subtract the cases corresponding to the first line of
 (\ref{trig82}) and put the common minus sign.

\section{Relation to the quantum-classical correspondence}

We have established the correspondence between solutions to the KZ equations
in different weight subspaces of $V^{\otimes n}$
and solutions to the spectral problem for the $n$-body Calogero model. It extends the
previously known Matsuo-Cherednik map to the case when $\mbox{dim}\, V$ is not necessarily
equal to $n$. In this more general form, the correspondence can be understood
as a natural ``quantization'' of the quantum-classsical correspondence
\cite{ALTZ14,GZZ14,TZZ15,MTV12} between
the quantum Gaudin model and the classical Calogero-Moser system of particles.

The Hamiltonian of the latter has the form
$$
{\cal H}=\sum_{i=1}^n p_i^2 -\sum_{i\neq j}^n\frac{\kappa^2}{(x_i-x_j)^2}
$$
with the usual Poisson brackets $\{p_i, x_j\}=\delta_{ij}$
(for simplicity we consider the rational case).
The model is known to be integrable \cite{Moser}, with the Lax matrix
$$
L_{ij}=p_i \delta_{ij}+\frac{\kappa (1-\delta_{ij})}{x_i-x_j}.
$$
The higher Hamiltonians in involution are given by traces of powers of the
Lax matrix:
$
{\cal H}_k =\mbox{tr}\, L^k$, ${\cal H}_2={\cal H}
$.
The correspondence with the quantum Gaudin model goes as follows.
Consider the level set of all classical Hamiltonians,
$$
{\cal H}_k =\sum_{a=1}^N M_a g_a^k, \qquad M_a\in \ZZ_{\geq 0},
$$
with fixed coordinates $x_i$. (This means that eigenvalues of the $n\! \times \! n$ Lax matrix
are $g_a$ with multiplicities $M_a$.)
Then the admissible values of momenta, $p_i$, coincide with
eigenvalues of the Gaudin Hamiltonians ${\bf H}_i$ in the weight subspace
${\cal V}(\{M_a\})$ for the model
with the marked points $x_i$ and the twist matrix
${\bf g}=\mbox{diag}\, (g_1, \ldots , g_N)$. In fact the admissible values of $p_i$'s obey
a system of algebraic equations. Different solutions of this system correspond to
different eigenstates of the Gaudin Hamiltonians.
The coupling constant $\kappa$ plays the role of the
formal Planck constant in the Gaudin model.

In the trigonometric case eigenvalues of the Lax matrix
$$
L_{ij}^{\rm trig}=p_i \delta_{ij}+\frac{\kappa \gamma
(1-\delta_{ij})}{\sinh \gamma (x_i-x_j)}
$$
should form
``strings'' of lengths $M_a$ centered at $g_a$ (see \cite{BLZZ16}):
$$
g_a^{(\alpha )}=g_a - (M_a-1-2\alpha )\kappa \gamma , \qquad \alpha =0,1, \ldots , M_a-1.
$$
Then $p_i$ are eigenvalues of the trigonometric Gaudin Hamiltonians.
The formula (\ref{trig3}) for the eigenvalue $E$ of the Calogero-Sutherland Hamiltonian
agrees with this since it is actually equal to the sum of squares of all $n$
eigenvalues of the trigonometric Lax matrix:
$$
E=\sum_{a=1}^{N}\sum_{\alpha =0}^{M_a-1}(g_a^{(\alpha )})^2,
$$
as one can easily check. Again, we conjecture that the function $\Psi$ is a common
eigenfunction for all higher Calogero-Sutherland Hamiltonians with eigenvalues
$\displaystyle{\sum_{a=1}^{N}\sum_{\alpha =0}^{M_a-1}(g_a^{(\alpha )})^k}$.

We see that the quantization of the classical Calogero system of particles
with the Planck constant $\hbar$
($p_i \to \hbar \p_{x_i}$) corresponds to the non-autonomous deformation
of the Gaudin model which is the system of KZ equations with the twist matrix.

\section{Conclusion}

In this paper we have discussed the Matsuo-Cherednik type correspondence between
solutions to the rational or trigonometric 
KZ equations in $(\CC^N)^{\otimes n}$ and solutions to the spectral
problem for the $n$-body Calogero model (respectively, rational or trigonometric). 
The previously known construction 
\cite{Matsuo,Cherednik} is extended to the case when $n$ is not necessarily equal to $N$.
The wave function of the Calogero model is simply a sum of all components
of a solution to the KZ equation in a given weight subspace. We also conjecture that
this wave function is a common eigenfunction for all higher commuting Calogero Hamiltonians.
This is checked by a direct calculation for the third (cubic) Calogero Hamiltonian.

It is important to note that this result sheds some new light on the 
quantum-classical correspondence between the quantum 
Gaudin model and the classical Calogero system of particles \cite{ALTZ14,GZZ14,TZZ15,MTV12}. 
Namely, it suggests what happens 
with the other side of the correspondence when the Calogero system gets quantized:
the Gaudin spectral problem should be substituted by its non-stationary version which is 
just the system of KZ equations.

\section*{Appendix: the cubic Hamiltonian}
In this appendix we show that the wave function $\Psi$ (\ref{kz2})
is an eigenfunction of the third (cubic) Calogero Hamiltonian
$$
\hat {\cal H}_3=\sum_i \hbar^3
\p_{x_i}^3- 3\hbar\ka(\ka-\hbar)\sum_{i\neq j}
\frac{1}{(x_i-x_j)^2}\,\p_{x_i}.
$$
 It is convenient to introduce the KZ connection
 \beq\label{cu1}
  \nabla_i=\hbar \p_{x_i}-{\bf g}^{(i)}- \ka \sum_{j\neq i}^n \frac{{\bf
  P}_{ij}}{x_i-x_j}\,.
 \eeq
Then the KZ equations (\ref{kz1}) are of the form:
 \beq\label{cu2}
  \nabla_i\Bigl | \Phi \Bigr >=0\,,\quad\quad i=1,..,n\,.
 \eeq
Below we will omit the subscript $j\neq i$ implying that all
summation indices do not equal to $i$. Direct
calculations yield
  \beq\label{cu3}
  \begin{array}{l}
 \displaystyle{  \nabla_i^2=\hbar^2 \p_{x_i}^2-2\hbar\left( {\bf  g}^{(i)}+\ka\sum_j \frac{{\bf  P}_{ij}}{x_i-x_j}
 \right)\p_{x_i}+\left( {\bf  g}^{(i)}\right)^2+\ka\sum_j
 \frac{ {\bf  g}^{(i)}{\bf  P}_{ij}+{\bf  P}_{ij}{\bf  g}^{(i)} }{x_i-x_j}
  }
  \\ \ \\
 \displaystyle{ +\, \, \hbar\ka\sum_j \frac{{\bf  P}_{ij}}{(x_i-x_j)^2} +\ka^2\sum_{j,k}
 \frac{ {\bf P}_{ij}{\bf P}_{ik} }{ (x_i-x_j)(x_i-x_k)
 }\,,
  }
  \end{array}
 \eeq

  \beq\label{cu4}
  \begin{array}{l}
 \displaystyle{  \nabla_i^3=\hbar^3 \p_{x_i}^3-3\hbar^2\left( {\bf  g}^{(i)}+\ka\sum_j \frac{{\bf  P}_{ij}}{x_i-x_j}
 \right)\p_{x_i}^2 + 3\hbar^2\ka\sum_j \frac{{\bf
 P}_{ij}}{(x_i-x_j)^2}\,\p_{x_i} -\left( {\bf  g}^{(i)}\right)^3
  }
  \\ \ \\
 \displaystyle{ + 3\hbar\ka^2\sum_{j,k} \frac{ {\bf P}_{ij}{\bf P}_{ik} }{ (x_i-x_j)(x_i-x_k)}\,\p_{x_i}
  +3\hbar\left( {\bf  g}^{(i)}\right)^2\,\p_{x_i}
  +3\hbar\ka\sum_j \frac{ {\bf  g}^{(i)}{\bf  P}_{ij}+{\bf  P}_{ij}{\bf  g}^{(i)} }{x_i-x_j}\,\p_{x_i}
 }
   \\ \ \\
 \displaystyle{-2\hbar^2\ka\sum_j \frac{{\bf  P}_{ij}}{(x_i-x_j)^3}
 -\hbar\ka\sum_j \frac{ 2{\bf  g}^{(i)}{\bf  P}_{ij}+{\bf  P}_{ij}{\bf  g}^{(i)} }{(x_i-x_j)^2}
 -3\hbar\ka^2\sum_{j,k} \frac{ {\bf P}_{ij}{\bf P}_{ik} }{ (x_i-x_j)^2(x_i-x_k)}
 }
    \\ \ \\
 \displaystyle{-\ka\sum_j \frac{ \left({\bf  g}^{(i)}\right)^2{\bf  P}_{ij}+{\bf  g}^{(i)}{\bf  P}_{ij}{\bf  g}^{(i)}
  +{\bf  P}_{ij}\left({\bf  g}^{(i)}\right)^2}{(x_i-x_j)}
-\ka^3\sum_{j,k,l} \frac{ {\bf P}_{ij}{\bf P}_{ik}{\bf P}_{il}
}{(x_i-x_j)(x_i-x_k)(x_i-x_l)}
 }
    \\ \ \\
 \displaystyle{-\ka^2\sum_{j,k} \frac{ {\bf  g}^{(i)}{\bf P}_{ij}{\bf P}_{ik}+{\bf P}_{ij}{\bf  g}^{(i)}{\bf P}_{ik}+
  {\bf P}_{ij}{\bf P}_{ik}{\bf  g}^{(i)} }{ (x_i-x_j)(x_i-x_k)}\,.
 }
  \end{array}
 \eeq
Now let us use (\ref{cu2}). Substitute $\p_{x_i}^2$ from equation
$\nabla_i^2\Bigl | \Phi \Bigr >=0$ with $\nabla_i^2$ written as in
(\ref{cu3}):
  \beq\label{cu5}
  \begin{array}{l}
 \displaystyle{  \nabla_i^3=\hbar^3 \p_{x_i}^3 + 3\hbar^2\ka\sum_j \frac{{\bf
 P}_{ij}}{(x_i-x_j)^2}\,\p_{x_i} +2\ka^3\sum_{j,k,l} \frac{ {\bf P}_{ij}{\bf P}_{ik}{\bf P}_{il}
}{(x_i-x_j)(x_i-x_k)(x_i-x_l)}
  }
  \\ \ \\
 \displaystyle{ - 3\hbar\ka^2\sum_{j,k} \frac{ {\bf P}_{ij}{\bf P}_{ik} }{ (x_i-x_j)(x_i-x_k)}\,\p_{x_i}
  -3\hbar\left( {\bf  g}^{(i)}\right)^2\,\p_{x_i}
  -3\hbar\ka\sum_j \frac{ {\bf  g}^{(i)}{\bf  P}_{ij}+{\bf  P}_{ij}{\bf  g}^{(i)} }{x_i-x_j}\,\p_{x_i}
 }
   \\ \ \\
 \displaystyle{
 +\hbar\ka\sum_j \frac{ {\bf  g}^{(i)}{\bf  P}_{ij}-{\bf  P}_{ij}{\bf  g}^{(i)} }{(x_i-x_j)^2}
+2\ka^2\sum_{j,k} \frac{ {\bf  g}^{(i)}{\bf P}_{ij}{\bf P}_{ik}+{\bf
P}_{ij}{\bf  g}^{(i)}{\bf P}_{ik}+
  {\bf P}_{ij}{\bf P}_{ik}{\bf  g}^{(i)} }{ (x_i-x_j)(x_i-x_k)}
 }
    \\ \ \\
 \displaystyle{+2\ka\sum_j \frac{ \left({\bf  g}^{(i)}\right)^2{\bf  P}_{ij}+{\bf  g}^{(i)}{\bf  P}_{ij}{\bf  g}^{(i)}
  +{\bf  P}_{ij}\left({\bf  g}^{(i)}\right)^2}{(x_i-x_j)}-2\hbar^2\ka\sum_j \frac{{\bf  P}_{ij}}{(x_i-x_j)^3}
+2\left( {\bf  g}^{(i)}\right)^3\,.
 }
  \end{array}
 \eeq
 Here and below we imply that the operators act to a solution $\Bigl |\Phi \Bigr >$ of the
 KZ equation (we do not write the vector $\Bigl |\Phi \Bigr >$ for brevity).
In the same way make the following substitutions into the r.h.s. of
(\ref{cu5}) (using $\nabla_i\Bigl | \Phi \Bigr >=0$):
 $$
-3\hbar\left( {\bf  g}^{(i)}\right)^2\,\p_{x_i}=-3\hbar\left( {\bf
g}^{(i)}\right)^3-3\ka\sum_j \frac{ \left({\bf g}^{(i)}\right)^2{\bf
P}_{ij} }{(x_i-x_j)}\,,
 $$
$$
-3\hbar\ka\sum_j \frac{ {\bf  g}^{(i)}{\bf  P}_{ij}+{\bf P}_{ij}{\bf
g}^{(i)} }{x_i-x_j}\,\p_{x_i}
$$
$$
=-3\ka^2\sum_{j,k} \frac{ {\bf  g}^{(i)}{\bf P}_{ij}{\bf
P}_{ik}+{\bf P}_{ij}{\bf  g}^{(i)}{\bf P}_{ik} }{
(x_i-x_j)(x_i-x_k)}
  -3\ka\sum_j \frac{ {\bf
g}^{(i)}{\bf  P}_{ij}{\bf  g}^{(i)}
  +{\bf  P}_{ij}\left({\bf  g}^{(i)}\right)^2}{(x_i-x_j)}\,.
$$
Then we get
  \beq\label{cu6}
  \begin{array}{l}
 \displaystyle{  \nabla_i^3=\hbar^3 \p_{x_i}^3 + 3\hbar^2\ka\sum_j \frac{{\bf
 P}_{ij}}{(x_i-x_j)^2}\,\p_{x_i} +2\ka^3\sum_{j,k,l} \frac{ {\bf P}_{ij}{\bf P}_{ik}{\bf P}_{il}
}{(x_i-x_j)(x_i-x_k)(x_i-x_l)}
  }
   \\ \ \\
 \displaystyle{
 +\hbar\ka\sum_j \frac{ {\bf  g}^{(i)}{\bf  P}_{ij}-{\bf  P}_{ij}{\bf  g}^{(i)} }{(x_i-x_j)^2}
+\ka^2\sum_{j,k} \frac{ -{\bf  g}^{(i)}{\bf P}_{ij}{\bf P}_{ik}-{\bf
P}_{ij}{\bf  g}^{(i)}{\bf P}_{ik}+
  2{\bf P}_{ij}{\bf P}_{ik}{\bf  g}^{(i)} }{ (x_i-x_j)(x_i-x_k)}
 }
    \\ \ \\
 \displaystyle{+2\ka\sum_j \frac{ \left({\bf  g}^{(i)}\right)^2{\bf  P}_{ij}+{\bf  g}^{(i)}{\bf  P}_{ij}{\bf  g}^{(i)}
  +{\bf  P}_{ij}\left({\bf  g}^{(i)}\right)^2}{(x_i-x_j)}-2\hbar^2\ka\sum_j \frac{{\bf  P}_{ij}}{(x_i-x_j)^3}
 -\left( {\bf  g}^{(i)}\right)^3
 }
   \\ \ \\
 \displaystyle{ - 3\hbar\ka^2\sum_{j,k} \frac{ {\bf P}_{ij}{\bf P}_{ik} }{
 (x_i-x_j)(x_i-x_k)}\,\p_{x_i}\,.

 }
  \end{array}
 \eeq
The two sums over $j,k$ in (\ref{cu6}) should be subdivided into two
parts each -- with $j=k$ and $j\neq k$. Then the last sum in
(\ref{cu6}) with $j\neq k$ should be transformed via $\nabla_i\Bigl
| \Phi \Bigr >=0$. This yields
  \beq\label{cu7}
  \begin{array}{l}
 \displaystyle{  \nabla_i^3=\hbar^3 \p_{x_i}^3 - 3\hbar\ka\sum_j \frac{\ka-\hbar{\bf
 P}_{ij}}{(x_i-x_j)^2}\,\p_{x_i} +2\ka^3\sum_{j,k,l} \frac{ {\bf P}_{ij}{\bf P}_{ik}{\bf P}_{il}
}{(x_i-x_j)(x_i-x_k)(x_i-x_l)}
  }
   \\ \ \\
 \displaystyle{
 +\hbar\ka\sum_j \frac{ {\bf  g}^{(i)}{\bf  P}_{ij}-{\bf  P}_{ij}{\bf  g}^{(i)} }{(x_i-x_j)^2}
-\ka^2\sum_{j\neq k} \frac{ {\bf  g}^{(i)}{\bf P}_{ij}{\bf
P}_{ik}+{\bf P}_{ij}{\bf  g}^{(i)}{\bf P}_{ik}+
  {\bf P}_{ij}{\bf P}_{ik}{\bf  g}^{(i)} }{ (x_i-x_j)(x_i-x_k)}
 }
    \\ \ \\
 \displaystyle{+2\ka\sum_j \frac{ \left({\bf  g}^{(i)}\right)^2{\bf  P}_{ij}+{\bf  g}^{(i)}{\bf  P}_{ij}{\bf  g}^{(i)}
  +{\bf  P}_{ij}\left({\bf  g}^{(i)}\right)^2}{(x_i-x_j)}-2\hbar^2\ka\sum_j \frac{{\bf  P}_{ij}}{(x_i-x_j)^3}
 -\left( {\bf  g}^{(i)}\right)^3
 }
   \\ \ \\
 \displaystyle{ +\ka^2\sum_j \frac{{\bf  g}^{(i)}-{\bf  g}^{(j)}}{(x_i-x_j)^2}
-3\ka^3\sum_{j\neq k,l} \frac{ {\bf P}_{ij}{\bf P}_{ik}{\bf P}_{il}
}{(x_i-x_j)(x_i-x_k)(x_i-x_l)}\,.
 }
  \end{array}
 \eeq
 At last notice that in two sums over three indices $j,k,l$ the
 terms
 corresponding to coinciding indices cancel out. Finally, we
 have
  \beq\label{cu8}
  \begin{array}{l}
 \displaystyle{  \nabla_i^3=\hbar^3 \p_{x_i}^3 - 3\hbar\ka\sum_j \frac{\ka-\hbar{\bf
 P}_{ij}}{(x_i-x_j)^2}\,\p_{x_i} -\ka^3\sum_{j,k,l}{}^{'} \frac{ {\bf P}_{ij}{\bf P}_{ik}{\bf P}_{il}
}{(x_i-x_j)(x_i-x_k)(x_i-x_l)}
  }
   \\ \ \\
 \displaystyle{
 -\ka\sum_j \frac{ (\ka-\hbar{\bf
 P}_{ij}) ({\bf  g}^{(j)}-{\bf  g}^{(i)}) }{(x_i-x_j)^2}
-\ka^2\sum_{j\neq k} \frac{ {\bf P}_{ij}{\bf P}_{ik}\left[{\bf
g}^{(i)}+{\bf  g}^{(j)}+ {\bf  g}^{(k)}\right] }{
(x_i-x_j)(x_i-x_k)}
 }
    \\ \ \\
 \displaystyle{+2\ka\sum_j \frac{ {\bf  P}_{ij}\left[\left({\bf  g}^{(i)}\right)^2+{\bf  g}^{(i)}{\bf  g}^{(j)}
  +\left({\bf  g}^{(j)}\right)^2\right]}{(x_i-x_j)}-2\hbar^2\ka\sum_j \frac{{\bf  P}_{ij}}{(x_i-x_j)^3}
 -\left( {\bf  g}^{(i)}\right)^3\,,
 }
  \end{array}
 \eeq
where $\displaystyle{\sum_{j,k,l}{}^{'}}$ denotes summation over all distinct indices. Now we can
write:
  \beq\label{cu9}
  \begin{array}{l}
 \displaystyle{
\sum_i\Bigl<\Omega\Bigr |\nabla_i^3  \Bigl | \Phi \Bigr >=0=
 }
 \\ \ \\
 \displaystyle{
=\sum_i \hbar^3 \p_{x_i}^3\Psi - 3\hbar\ka(\ka-\hbar)\sum_{i\neq j}
\frac{1}{(x_i-x_j)^2}\,\p_{x_i}\Psi -\sum_i\Bigl<\Omega\Bigr
|\left({\bf g}^{(i)}\right)^3 \Bigl | \Phi \Bigr >
 }
  \end{array}
 \eeq
or
  \beq\label{cu11}
  \begin{array}{l}
 \displaystyle{
\hat {\cal H}_3\Psi=E\Psi\,,\quad \hat {\cal H}_3=\sum_i \hbar^3
\p_{x_i}^3- 3\hbar\ka(\ka-\hbar)\sum_{i\neq j}
\frac{1}{(x_i-x_j)^2}\,\p_{x_i}
 }
  \end{array}
 \eeq
where
  \beq\label{cu12}
  \begin{array}{l}
 \displaystyle{
E=\sum\limits_{a=1}^N M_a g_a^3\,.
 }
  \end{array}
 \eeq
In transition from (\ref{cu8}) to (\ref{cu9}) we have used
$\Bigl<\Omega\Bigr| {\bf  P}_{ij}=\Bigl<\Omega\Bigr|$ and the identities
  \beq\label{cu13}
  \begin{array}{l}
 \displaystyle{
\sum_{i,j,k}{}^{'} \frac{ {\bf g}^{(i)}+{\bf  g}^{(j)}+ {\bf g}^{(k)}
}{ (x_i-x_j)(x_i-x_k)}=0\,,
 }
  \end{array}
 \eeq
  \beq\label{cu14}
  \begin{array}{l}
 \displaystyle{
\sum_{i,j,k,l}{}^{'} \frac{ 1 }{(x_i-x_j)(x_i-x_k)(x_i-x_l)}=0\,.
 }
  \end{array}
 \eeq
The other terms cancel due to skew-symmetry with respect to
$i,j$.

\section*{Acknowledgments}

The work of A. Zabrodin has been funded by the Russian Academic
Excellence Project `5-100'. It was also supported in part by RFBR
grant 15-01-05990 and by joint RFBR grant 15-52-50041 YaF$_a$. The
work of A. Zotov was supported by RFBR grant 15-31-20484
mol$\_$a$\_$ved and by joint RFBR project 15-51-52031 HHC$_a$.

\end{document}